\documentclass[aps,pre,preprint,longbibliography]{revtex4-2}

\usepackage{amsmath}
\usepackage{amssymb}
\usepackage{graphicx}
\usepackage[svgnames]{xcolor}
\usepackage[colorlinks=true,citecolor=blue,linkcolor=blue,citebordercolor=white,linkbordercolor=white]{hyperref}

\def\eq#1{{Eq.~(\ref{#1})}}

\def\CURRENT_END{\bigskip
\noindent{\begin{center}\color{red}\rule{0.25\textwidth}{8pt}
\hskip7pt \textcolor{DarkRed}{\textsc{End for now: \today}} \hskip7pt \color{red}\rule{0.25\textwidth}{8pt}\end{center}}
\bigskip}

\begin{document}

\title[Charge transport in the spatially correlated exponential random energy
landscape]{Charge transport in the spatially correlated exponential random energy
landscape: effect of the non-positive correlation function}

\author{S.V. Novikov}
\email{novikov@elchem.ac.ru}
\affiliation{A.N. Frumkin Institute of
Physical Chemistry and Electrochemistry, Leninsky prosp. 31,
Moscow 119071, Russia}
\affiliation{National Research University Higher School of Economics, Myasnitskaya Ulitsa 20, Moscow 101000, Russia}


\date{\today}

\begin{abstract}
Charge transport in amorphous semiconductors having spatially correlated exponential density of states (DOS) has been considered for the arbitrary behavior of the correlation function of random energies. Average carrier velocity is exactly calculated for the quasi-equilibrium (nondispersive) transport regime.
For the symmetric exponential DOS with exponential tails for low and high energies and non-positive correlation function the temperature for the transition to the dispersive transport regime depends on correlation properties and becomes greater than the traditional estimation based on the DOS decay energy $kT=U_0$.
Another new feature of the transport in the landscape having non-positive correlation function is the decay of the mobility with field in low field region.
\end{abstract}

\maketitle

\section{Introduction}

Major characteristics of a random medium affecting the hopping transport of charges in amorphous semiconductors are the density of states and spatial correlation of the random energy landscape. \cite{pollak1991hopping,pope1999,Dunlap:542,Protopopova:9020,kohler2015electronic,Baranovskii:487}
Here we consider highly disordered semiconductors having disorder high enough to provide a full localization of all relevant states. The most prominent effect of the spatial correlation is its dominating role in the formation of the particular mobility field dependence $\mu(E)$. This statement can be illustrated by the best studied case of the Gaussian DOS; such DOS is frequently used to describe charge transport in amorphous organic semiconductors.\cite{Bassler:15,Dieckmann:8136} In the simplest model of charge transport (continuous diffusive model)  we assume that the particle moves in the random energy landscape $U(x)$ under the action of the applied electric field $E$ and the behavior of the particle density $n(x,t)$ is governed by the equation
\begin{equation}\label{diffusion1}
\frac{\partial n}{\partial t}=D_0 \frac{\partial}{\partial x}\left[\frac{\partial n}{\partial x}+\left(\frac{1}{kT}\frac{\partial U}
{\partial x}-\gamma \right)n\right], \hskip10pt \gamma=\frac{v_0}{D_0},
\end{equation}
where $v_0\propto E$ and $D_0$ are bare velocity and diffusivity, correspondingly. In the continuous model we assume that the relevant distances are much larger than the typical hopping distance and are overcome by many hops. Long time and long range features of the hopping transport are well described by this approximation. For the Gaussian DOS the resulting expression for the average carrier velocity is
\begin{equation}
v=\mu E=\frac{D_0}{\int_0^\infty dx \exp\left\{-e\beta E
x+\beta^2\sigma^2\left[c(0)-c(x)\right]\right\}}, \label{1D_mu}
\end{equation}
where $\beta=1/kT$, $\sigma$ is the rms disorder, and correlation function $c(x)=\left<U(x)U(0)\right>/\sigma^2$ is normalized as $c(0)=1$ (we assume here that $\left<U\right>=0$, angular brackets mean an average over realizations of disorder).\cite{Dunlap:542}

From \eq{1D_mu} immediately follows that if $c(x)\propto 1/x^n$ for $x\rightarrow \infty$ then the leading asymptotics is
\begin{equation}\label{power}
  \ln\mu\propto E^{n/(n+1)}
\end{equation}
for moderate $E$ and strong disorder $\sigma\beta \gg 1$.\cite{Dunlap:542} Power law correlation functions naturally emerge in reasonable models of the amorphous organic semiconductors with dominating electrostatic disorder. In organic materials we have a negligible concentration of intrinsic free carriers and the lack of screening, thus dipoles or quadrupoles provide long range contributions to the energy landscape. Long range contributions inevitably lead to the correlated Gaussian landscape with the power law correlation functions having exponent $n=1$ for dipoles and $n=3$ for quadrupoles.\cite{Novikov:14573,Dunlap:542,Novikov:181}

Our confidence in the existence of the strong spatial correlation of random energies in amorphous organic semiconductors is based on the reasonable explanation of the frequent occurence of the so called Poole-Frenkel mobility field dependence $\ln\mu\propto E^{1/2}$ in polar organic materials with dominant dipolar disorder.\cite{Dunlap:542} Dominance of the dipolar disorder and the proper behavior of the correlation function are supported by the large-scale computer simulation of the structure of typical amorphous organic semiconductors.\cite{Masse:115204} Computer simulation shows that for 3D charge transport the resulting dependence $\mu(T,E)$ differ from the corresponding 1D dependence by the change of some numerical coefficients.\cite{Novikov:4472, Novikov:954}

Nonetheless, effects of spatial correlation on charge transport are still underestimated and in many recent studies it is assumed that the random energy landscape in amorphous organic materials is uncorrelated.\cite{Melianas:1806004,Kim:106402} To a major extent this is motivated by the comparative simplicity of the uncorrelated landscape and absence of the direct experimental data on the correlation properties of real amorphous materials.  Additionally, the case of very short range correlation in many situations can be adequately described by the uncorrelated picture. Still, the correlated random  energy landscape provides the most consistent explanation of the development of the Pool-Frenkel mobility field dependence in organic semiconductors.

Another popular model of the DOS in amorphous materials is the exponential DOS (eDOS), i.e. the density with the exponential tail
\begin{equation}\label{exp-DOS}
p(U)=\frac{1}{U_0}e^{U/U_0},\hskip10pt U < 0.
\end{equation}
Tail states are the most important for the description of the long-time transport properties. Specific shape of eDOS is responsible for  the development of the highly non-equilibrium transport for the low temperature $kT < U_0$ with the power-law temporal dependence of the current transients, both for the multiple-trapping and hopping models of charge transport. \cite{Shlesinger:421,Rudenko:209,Hartenstein:8574}

Influence of the spatial correlation in eDOS on the transport properties is  poorly studied. For example, usually the consideration of the charge transport in amorphous organic semiconductors having exponential DOS does not take into account any possible correlation effect, \cite{Vissenberg:12964,Tachiya:85201,Street:165207,Schubert:24203} though in such materials the spatial correlation of the energy landscape naturally emerges. \cite{Novikov:14573,Dunlap:542,Novikov:2584}

Recently the behavior of the average carrier velocity $v$ and diffusivity $D$ for the hopping transport in the disordered medium having  spatially correlated eDOS has been studied for the continuous model of charge transport. \cite{Novikov:24504,Novikov:24505} The average particle velocity in the infinite transport medium is
\begin{equation}\label{stat-v_inf}
v=\frac{D_0}{\int\limits_0^\infty dx \hskip2pt\exp\left(-\gamma x\right)Z(x)}, \hskip10pt Z(x)=\left<\hskip2pt\exp\left[\frac{U(x)-U(0)}{kT}\right]\right>,
\end{equation}
where angular brackets mean statistical averaging over realization of $U(x)$. \cite{Parris:2803,Parris:5295,Kenkre:99}

Spatial correlation in $U(x)$ was introduced by representing the exponentially distributed random energy as a combination of the independent Gaussian random variables $X$, $Y$ having zero mean and unit variance
\begin{equation}
U=-\frac{1}{2}U_0\left(X^2+Y^2\right).
\label{U-XY_a}
\end{equation}
If $X(x),\hskip3pt Y(x)$ are spatially correlated random fields with correlation functions $c_X(x)$, $c_Y(x)$, then the random energy $U(x)$ is correlated, too, and the corresponding correlation function
\begin{equation}
c_U(x)=\left<U(x)U(0)\right>-\left<U\right>^2=\frac{1}{2}U^2_0\left[c_X(x)^2+c_Y^2(x)\right]
\label{cE-old}
\end{equation}
is strictly nonnegative. Representation (\ref{U-XY_a}) may have a physical justiﬁcation in some amorphous organic materials: contribution from the molecular polarizability describes the random energy as a sum of squares of three Gaussian variables and gives the density of states $p(U)\propto (-U)^{1/2}\exp(U/U_0)$, very close to the exponential one.\cite{May:136401}

The statistical average $Z(x)$ for the high temperature case $U_0/kT < 1$ was calculated for $c_X(x)=c_Y(x)= c(x)$
\begin{equation}
Z(x)=\frac{1}{1-\varkappa^2\left[1-c^2(x)\right]}, \hskip10pt \varkappa=U_0/kT
\label{Z}
\end{equation}
and then we calculated the mobility field dependence for various types of correlation functions. It was found that spatial correlation significantly changes transport properties, determines the dependence of the mobility on $v_0$, and provides unusual transport properties (breakdown of the normal diffusion) in the case of the long range correlations.\cite{Novikov:24504,Novikov:24505} If $Z(x)\rightarrow \infty$, then $v$ goes to $0$, thus signalling the transition to the dispersive transport regime where the average carrier velocity depends on the thickness $L$ of the transport layer. Divergence of $Z(x)$ takes place when
\begin{equation}\label{zero}
  1-\varkappa^2\left[1-c^2(x)\right] \rightarrow 0
\end{equation}
and the true critical value $\varkappa_c$ is defined by the minimal value of $c^2(x)$, i.e. $c(x)=0$ which is inevitably achieved for $x\rightarrow \infty$. Hence, the transition temperature $kT_c=U_0$ does not depend on the correlation and coincides with the temperature of the breakdown of the quasi-equilibrium density of the occupied states
\begin{equation}\label{occDOS}
  p_{occ}(U)\propto p(U)\exp\left(-U/kT\right)\propto \exp\left(\frac{U}{U_0}-\frac{U}{kT}\right).
\end{equation}
For $kT < U_0$ the density $p_{occ}(U)$ becomes non-normalizable.

Symmetric Gaussian-generated DOS (sDOS)
\begin{equation}\label{sexp}
p_s(U)=\frac{1}{2U_0}\exp\left(-\frac{|U|}{U_0}\right), \hskip10pt -\infty < U <\infty
\end{equation}
could be represented as
\begin{equation}\label{sdosG}
  U=\frac{1}{2}U_0\left(X_1^2+Y_1^2-X_2^2-Y_2^2\right),
\end{equation}
where $X_i$, $Y_i$ are again independent zero-average Gaussian variables with the rms disorder $\sigma=1$. For such representation the correlation function $c_U(x)$ is again nonnegative, thus most transport properties are very close to the corresponding properties of the eDOS. For example, the function $Z(x)$ has the form
\begin{equation}\label{Zs}
  Z(x)=\frac{1}{\left(1-\varkappa^2\left[1-c^2(x)\right]\right)^2},
\end{equation}
and, again, for the Gaussian-induced sDOS (\ref{sdosG}) transition to the dispersive regime takes place at $kT_c=U_0$  and does not depend on the correlation properties.

Obviously, the transport behavior for the case where $c_U(x)$ could become negative for some $x$ is not captured by the representation (\ref{U-XY_a}) or (\ref{sdosG}). It should be noted that the Gaussian representations (\ref{U-XY_a}) and (\ref{sdosG}) certainly impose other limitation to statistical properties of $U(x)$, thus raising a natural question: which transport properties are in principal possible for the correlated exponential DOS (eDOS or sDOS) but not captured by the representation (\ref{U-XY_a}) or (\ref{sdosG})? May we expect some really unusual effects for the nonpositive $c_U(x)$?

Nonpositive correlation functions do occur in many areas of physics, but for our study the most important example is the nonpositive energy-energy correlation function in the particular model of the disordered material with the dominant electrostatic contribution to the total energetic disorder (sources of the disorder are induced dipoles).\cite{Gartstein:351,Freire:134901} For this particular case it was found that the spatial disorder (when sources of the disorder are not located on the sites of a regular lattice) facilitates the development of the negative correlation.\cite{Freire:134901} Even more simple model with the nonpositive correlation function is the model of randomly oriented permanent dipoles arranged on 1D line while for 2D and 3D cases the corresponding functions are positive.\cite{Novikov:14573} We may suppose that if we consider the disorder developed from the contribution of many uncorrelated sources, then the low dimensionality, spatial disorder in the location of sources, and not very slow decay of the contribution from the individual source are favorable for the development of the negative correlation.

\begin{figure}[tbp]
\centering
\includegraphics[width=3.375in]{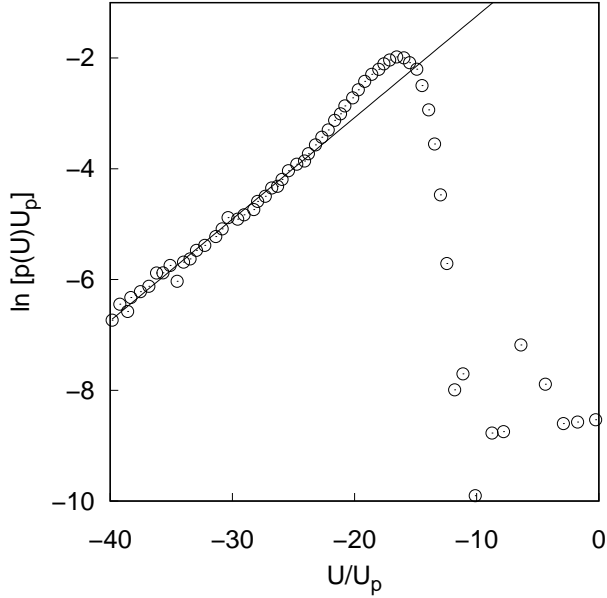}
\caption{Density of states for the model considered in ref. \citenum{Freire:134901} (data is borrowed from Fig. 1b). Random displacement vectors were taken from the Gaussian distribution with the rms disorder equal to $0.3a$ where $a$ is the lattice constant, and sites were occupied by molecules having polarizability $\alpha$. Scale energy $U_p$ is equal to $e^2\alpha/2a^4$. Circles represent the simulation data and the line shows the best linear fit.}
\label{Tail}
\end{figure}

Moreover, for the case of moderate spatial disorder the DOS presented in ref. \citenum{Freire:134901} develops a clear exponential tail (see Fig. \ref{Tail}, this is the case where induced dipoles are located at the sites of a disordered lattice which was built by adding random displacement vectors to the sites of the simple cubic lattice). Hence, in this model we have a simultaneous development of the exponential DOS and nonpositive spatial correlation.

In this paper we are going to calculate the average velocity $v$ for the case of nonpositive $c_U(x)$ for the exponential DOS (\ref{exp-DOS}) and sDOS (\ref{sexp}) using the alternative, but relative approach to induce the spatial correlation. It turns out that the negative correlation significantly changes the transport properties. In addition, using the alternative representation for $U(x)$ for the positive $c_U(x)$ we can check the sensitivity of properties of the resulting exponential distribution to the type of the auxiliary Gaussian representation.

\section{Generation of the exponential random field with the nonpositive correlation function}

\begin{figure}[tbp]
\centering
\includegraphics[width=3.375in]{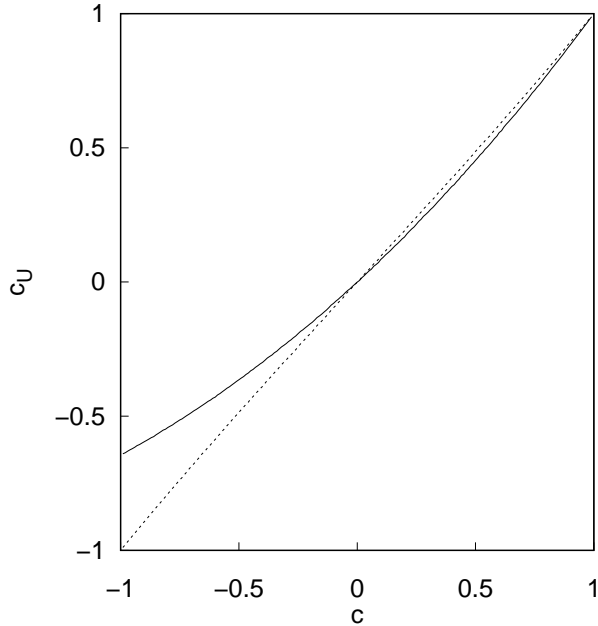}
\caption{Plot of the correlation coefficient $c_{U}$ as a function of the corresponding Gaussian coefficient $c$ for eDOS (solid line) and sDOS (broken line). For sDOS the dependence is almost indistinguishable from the straight line $c_{U}(c)=c$. }
\label{cG_cexp}
\end{figure}

We may derive the exponential density from the Gaussian one using a simple modification of the well-known inversion method.\cite{Devroye-book}  We can write
\begin{equation}
\exp\left(U/U_0\right)=\frac{1}{\sqrt{2\pi}}\int\limits_{-\infty}^X ds\exp\left(-s^2/2\right)=\Phi(X)=\frac{1}{2}\left[1+{\rm erf}\left(\frac{X}{\sqrt{2}}\right)\right],
\label{exp-from-G}
\end{equation}
and for the symmetric DOS the corresponding relation is
\begin{equation}
\exp\left(-|U|/U_0\right)=\begin{cases} 2 \Phi(X), \hskip10pt U < 0, \hskip10pt X < 0,\\ 2\left[1-\Phi(X)\right], \hskip10pt U >0, \hskip10pt X > 0. \end{cases}
\label{exp-from-Gs}
\end{equation}
If $X(x)$ is the spatially correlated random variable with the bivariate density
\begin{equation}\label{2a-PDF_GX}
p_G(X_1,X_2,c)=\frac{1}{2\pi\sqrt{1-c^2}}\exp\left[
-\frac{X_1^2+X_2^2-2c X_1 X_2}{2\left(1-c^2\right)}\right]
\end{equation}
then the  resulting exponential distribution of $U(x)$ is spatially correlated, too. Here $X_i=X(x_i)$ and $c=c(x_1-x_2)$ is the Gaussian correlation coefficient (or the correlation function, if we consider its dependence on $x_1-x_2$). Relation between $X(x)$ and $U(x)$ is more complicated in comparison with \eq{U-XY_a} but permits us to introduce the negative correlations for $U$ (see Fig. \ref{cG_cexp}; here and later we use the correlation function $c_U(x)$ for the exponential DOS normalized as $c_U(0)=1$). This approach may be considered as a particular example of the so-called Gaussian copula.\cite{balakrishnan2009}

 Correlation coefficient $c_{U}$ for sDOS is pretty close to the corresponding Gaussian coefficient $c$, while the dependence $c_{U}(c)$ for eDOS is nonlinear due to the asymmetry of the eDOS distribution (Fig. \ref{cG_cexp}). For this reason $c_U(1)=1$, while $c_U(-1)=1-\pi^2/6=-0.644...$ (this is the minimal possible value of $c_U$ for eDOS). Nonetheless, for $c > 0$ $c_{U}(c)\approx c$, unlike the case of the previously used Gaussian model (\ref{U-XY_a}) (XY-model). Formal minimal value of $c$ could be as low as $-1$ if we consider just some pair of random Gaussian variables $X_1$ and $X_2$, but for the random field $X(x)$ the situation becomes different. It is well known that the necessary and sufficient condition for $c(x)$ to be a proper correlation function is the nonnegativity of its Fourier transform.\cite{yaglom1987correlation}  For example, for the bi-exponential correlation function
\begin{equation}\label{exp-corr}
c(x)=A\exp(-x/a)+(1-A)\exp(-x/l)
\end{equation}
with weight $A >1$ and $l >a$ (so $c(x) < 0$ becomes possible for some $x$, this function will be used extensively as an example of the nonpositive correlation function), we must have
\begin{equation}\label{Acr}
A \le A_{c}=\frac{1}{1-s},\hskip10pt s=a/l <1.
\end{equation}
Simple analysis shows that the lowest $c_{min}$ for such class of functions is achieved for $A=A_{c}$ and $s\rightarrow 1$, where the limiting function is
\begin{equation}\label{limit}
c_{lim}(x)=\left(1-\frac{x}{a}\right)\exp\left(-\frac{x}{a}\right)
\end{equation}
and the corresponding $c_{min}=-1/e^2=-0.135...$. For other functional classes $c_{min}$ could be even lower (see later). Thus, though $c_{min}\simeq -1$ is probably not achievable, the actual values of $c_{min}\simeq -0.2$ are quite reasonable.

\section{Temperature of the transition to the dispersive transport}

The most salient feature of the exponential DOS is the transition to the dispersive transport regime when $v\rightarrow 0$ at some critical temperature $T_c$. At the critical temperature $Z(x)$ in \eq{stat-v_inf} goes to infinity. In the Gaussian representation this means the divergence of the integral at $|X_0|,|X| \rightarrow \infty$, here we use the Gaussian variable $X_0$ for the generation of $U(0)$ and $X$ for the generation of $U(x)$.

Let us consider the behavior of $Z(x)$ at any given $x$ and, hence, $c(x)$. Convergence of $Z$ is determined by large $X_0$ and $X$, where behavior of $U(X)$ becomes much simpler. For eDOS
\begin{equation}\label{Xminf}
U(X\rightarrow -\infty) = -\frac{U_0}{2} X^2+...
\end{equation}
and $U(X\rightarrow \infty)\rightarrow 0$. Obviously, the most dangerous case is $X_0\rightarrow -\infty$ and $X\rightarrow \infty$, where the difference $U(x)-U(0)$ grows. The quadratic form in the exponent of the integral for $Z$ becomes
\begin{equation}\label{S2}
S_2(X_0,X) = -\frac{X_0^2+X^2-2cX_0 X}{2(1-c^2)} +\frac{\varkappa}{2}X^2
\end{equation}
and it should be negatively defined for the convergence of the integral. Let us consider its behavior on the line $X_0=a X$, $a < 0$. For the negativity of $S_2$ we need
\begin{equation}\label{Sa}
\Sigma_2(a)= a^2+1-2ca- \varkappa (1-c^2) > 0.
\end{equation}
Inequality is invalid for
\begin{equation}\label{apm_e}
a_- < a < a_+, \hskip10pt a_\pm=c\pm \left[(\varkappa-1)(1-c^2)\right]^{1/2}.
\end{equation}
If $c < 0$, then the only necessary condition for the divergence is $\varkappa > 1$ (because $a$ must be negative), while for $c > 0$ we have
\begin{equation}\label{c>0}
(\varkappa-1)(1-c^2) > c^2, \hskip10pt \varkappa > \frac{1}{1-c^2}.
\end{equation}
Hence, finally for eDOS
\begin{equation}
\varkappa_c= \begin{cases} 1, \hskip10pt c < 0, \\ \frac{1}{1-c^2}, \hskip10pt c > 0. \end{cases}
\label{kappa_eDOS}
\end{equation}

For sDOS again the most dangerous case is $X_0\rightarrow -\infty$ and $X\rightarrow \infty$, where
\begin{equation}\label{sDOS-diff}
U(X)-U(X_0) = \frac{U_0}{2}\left( X_0^2+X^2\right)+...
\end{equation}
and the corresponding condition for the convergence is
\begin{equation}\label{sDOS-Sa}
a^2+1-2ca- \varkappa (1-c^2)(a^2+1) > 0
\end{equation}
with $a<0$ falling outside the interval between roots $a_-$, $a_+$
\begin{equation}\label{apm_s}
a_\pm=\frac{c\pm \left[c^2-\left(1-\varkappa(1-c^2)\right)^2\right]^{1/2}}{1-\varkappa(1-c^2)}.
\end{equation}
For $c > 0$ the only convergence condition is $1-\varkappa(1-c^2) > 0$, while for $c < 0$ the additional condition is $|c| <1-\varkappa(1-c^2)$, hence, finally
\begin{equation}
\varkappa_c= \begin{cases} \frac{1}{1+|c|}, \hskip10pt c < 0, \\ \frac{1}{1-c^2}, \hskip10pt c > 0. \end{cases}
\label{kappa_sDOS}
\end{equation}

Until now we consider the convergence of $Z$ at one particular value of $x$ and $c=c(x)$. If we consider the whole $x$ range then the true $\varkappa_c$ is determined by its lowest possible value. Generally, from the maximal value $c(0)=1$ correlation function decays (non-monotonously in the case of the nonpositive $c(x)$) to $0$ at the infinity. Hence, for eDOS the transition occurs at $\varkappa_c=1$ irrespectively of the behavior of $c(x)$, while for sDOS $\varkappa_c=1$ for the positively defined  $c(x)$, while for the nonpositive $c(x)$ the actual $\varkappa_c < 1$ is determined by the minimal value $c_{min} < 0$.

Hence, for sDOS with anti-correlation the transition to the unusual transport with $v=0$ occurs at higher temperature than the temperature of the breakdown of the stationary occupied DOS  which still occurs at $kT_b=U_0$. Anti-correlation leads to the increase of the effective barriers. Quite probably, the effect is strictly onedimensional and may be observed in the quasi-onedimensional charge transport in long polymer chains with the suitable DOS.

\section{More general case: asymmetric  exponential DOS}

\begin{figure}[htbp]
\centering
\includegraphics[width=3.375in]{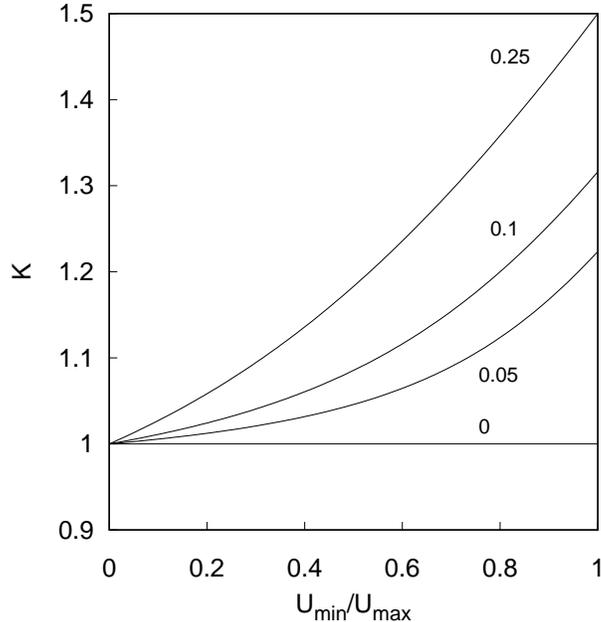}
\caption{Plot of the ratio $K=kT_c/U_{max}$ as a function of $U_{min}/U_{max}$ for several values of $c_{min}^2$ (indicated at the corresponding curve).}
\label{K}
\end{figure}

Let us consider the case where asymptotics of the DOS for $U\rightarrow -\infty$
\begin{equation}
\ln p(U)\simeq U/U_-
\label{-inf}
\end{equation}
and $U\rightarrow \infty$
\begin{equation}
\ln p(U)\simeq -U/U_+
\label{inf}
\end{equation}
are different. The analysis is identical to the one used in the previous section. The critical temperature for $c_{min}=0$ is again $kT_c=U_{max}=\max(U_-,U_+)$, while for $c_{min} < 0$ it is defined by the relation
\begin{equation}
kT_c=\frac{1}{2}\left(U_++U_-\right)\left(1+\sqrt{1-\rho}\right), \hskip10pt \rho=\frac{4(1-c_{min}^2)U_+U_-}{(U_++U_-)^2}.
\label{Tc}
\end{equation}
It is a symmetric function of $U_+$, $U_-$ and $kT_c > U_{max}$ for any $U_+ \ne U_-$ (Fig \ref{K}). For $U_+=U_-=U_0$ \eq{Tc} goes to \eq{kappa_sDOS} and it goes to  $kT_c=U_{max}$ for $U_{min}=\min(U_-,U_+)=0$ (one-sided distribution) or $c_{min} \rightarrow 0$.

The most interesting feature of the shift of $T_c$ for the anticorrelated sDOS is its dependence on the single parameter $c_{min}$. All other correlation characteristics (such as correlation length and functional form of the correlation function) are irrelevant. With this respect that property is similar to the behavior of the temperature of the breakdown of the quasi-equilibrium regime for the exponential DOS being again independent of any feature of the DOS apart from the decay parameter $U_0$.

\section{Mobility field dependence}

\begin{figure}[htbp]
\centering
\includegraphics[width=3.375in]{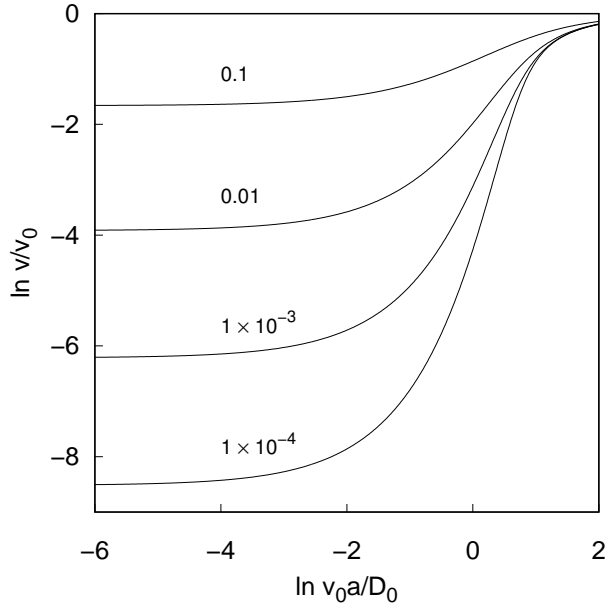}
\caption{Mobility field dependence for eDOS with the correlation function $c(x)=\exp(-x/a)$ for different values of the parameter $1-\varkappa$ (indicated near the corresponding curve). For sDOS the behavior is quite similar.}
\label{var-kappa}
\end{figure}

\begin{figure}[htbp]
\centering
\includegraphics[width=3.375in]{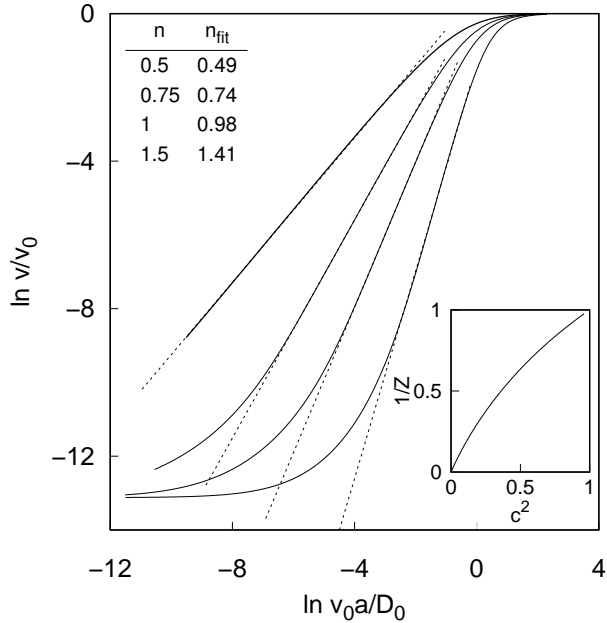}
\caption{Mobility field dependence (solid lines) calculated for eDOS and power law correlation function $c(x)=a^2/(x^2+a^2)^{n/2}$ for various values of $n$: 0.5, 0.75, 1, 1.5, from the upmost curve to the bottom, correspondingly;  $1-\varkappa= 1\times 10^{-6}$. Effective values of $n_{fit}$ for different $n$ are estimated from the best fit of the middle linear regions of the curves to the relation $v/v_0 \propto (\gamma a)^{2n}$ (broken lines). Inset shows the dependence of $1/Z$ on $c^2$ for $c>0$; for $c\rightarrow 0$ $1/Z \rightarrow \epsilon+{\rm const} \hskip2pt c^2$, and $\epsilon \rightarrow 0$ for $\varkappa\rightarrow 1$, exactly as in the case of XY-model.}
\label{power-law}
\end{figure}

For the discussion of the mobility field dependence we use the dependence of the dimensionless ratio $v/v_0=\mu/\mu_0$ on $v_0=\mu_0 E$ instead of $\mu(E)$, here $\mu_0$ is the mobility in the absence of disorder.  General behavior of the mobility field dependence demonstrates a lot of similarities with the mobility dependence in the XY-model.\cite{Novikov:24504}  For example, the field dependence for eDOS and sDOS becomes prominent only in the vicinity of the transition to the dispersive regime (Fig. \ref{var-kappa}). If $c(x) > 0$, then the functional form of the mobility field dependence is very close to the one obtained for the XY-model: for example, in the case of the power law correlation function $c(x)=a^n/(x^2+a^2)^{n/2}$ there is a power law mobility field dependence $v/v_0 \propto v_0^{2n}$, essentially the same as in the XY-model (see Fig. \ref{power-law}). The reason for the similarity is the same dependence of $Z$ on $c$ for $\varkappa\rightarrow 1$.

\begin{figure}[htbp]
\centering
\includegraphics[width=3.375in]{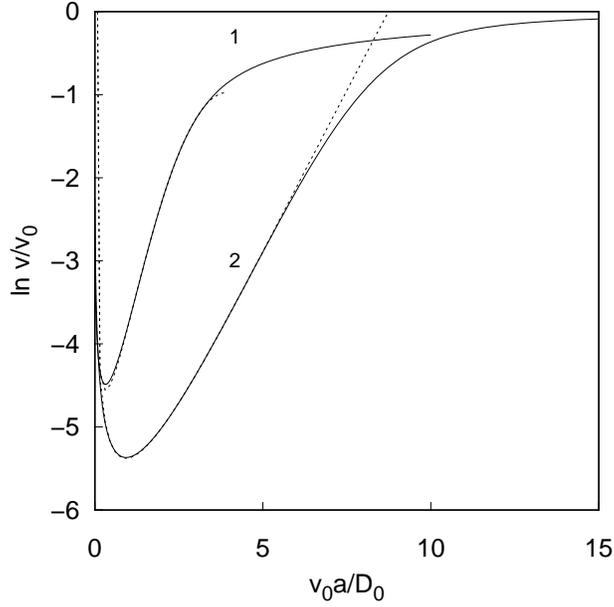}
\caption{Mobility field dependence for the sDOS with $c(x)=2\exp(-x/a)-\exp(-x/2a)$ (1, $\varkappa=0.88$ and $\varkappa_c=0.888...$) and  $c(x)=2.14 \frac{a^7}{(x^2+a^2)^{7/2}}-1.14 \frac{a^3}{(x^2+a^2)^{3/2}}$ (2, $\varkappa=0.82$ and $\varkappa_c=0.82446...$), solid lines. Broken lines show the best fit for \eq{vsp}, see explanation in the text. Weights $A$ for the correlation functions have been chosen to be very close to the corresponding critical values $A_c$ in order to minimize $c_{min}$. }
\label{dec_mu}
\end{figure}

A particular feature of the mobility field dependence for the  nonpositive correlation function $c(x)$ is the development of the low field region where the mobility decreases with $v_0$ (Fig. \ref{dec_mu}); this  behavior is impossible for the XY-model.\cite{Novikov:24504}  Another remarkable feature of the field dependence for the nonpositive $c(x)$ is the approximate universal linear field dependence of $\ln v$ for the moderate field strength.

\begin{figure}[htbp]
\centering
\includegraphics[width=3.375in]{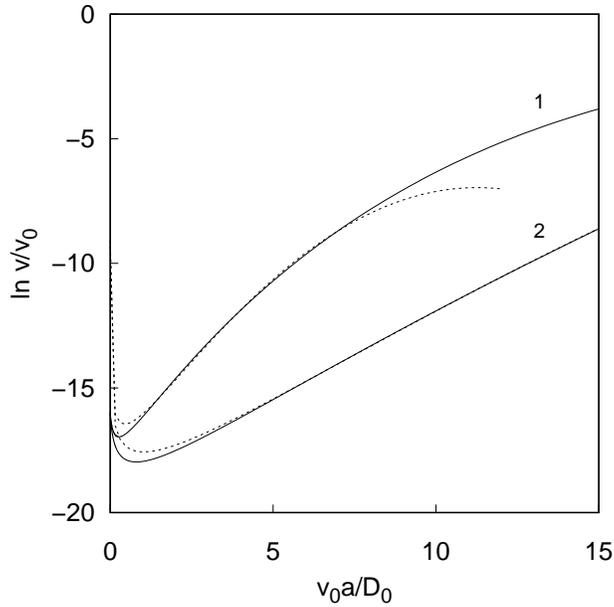}
\caption{The same plot as in Fig. \ref{dec_mu}  but for the Gaussian DOS with $\sigma\beta=4$.}
\label{Gauss}
\end{figure}

It is worth to note that the same kind of the field dependence takes place for the Gaussian DOS. For not so small $v_0$ the integral in \eq{1D_mu} is dominated by the negative minimum of $c(x)$. Assuming that in the vicinity of the minimum
\begin{equation}\label{x_{min}}
c(x) = c_{min} + \frac{(x-x_{min})^2}{2 p^2}+...,
\end{equation}
and $\sigma\beta \gg 1$, the saddle point method immediately gives
\begin{equation}\label{vsp}
\ln v/v_0\simeq -(\sigma\beta)^2 (1-c_{min})+\gamma x_{min}-\frac{(\gamma p)^2}{2(\sigma\beta)^2},
\end{equation}
we drop here the pre-exponential factor which does not depend on $\gamma$ for the parabolic dependence (\ref{x_{min}}) and the third term is the correction to the dominant linear contribution. Hence, the mobility field dependence has a universal functional form and does not depend on the details of $c(x)$ (see Fig. \ref{Gauss}). Fit of the calculated dependence for \eq{vsp} gives reasonable values for $x_{min}$, i.e. $x_{min}\approx 2.35a$ for  the exponential $c(x)$ and $x_{min}\approx 0.99a$ for the power law $c(x)$, while the exact values are $2.77a$ and $1.04a$, correspondingly. The agreement is quite good taking into account that the parabolic approximation (\ref{x_{min}}) differs from the exact correlation function $c(x)$.

\begin{figure}[htbp]
\centering
\includegraphics[width=3.375in]{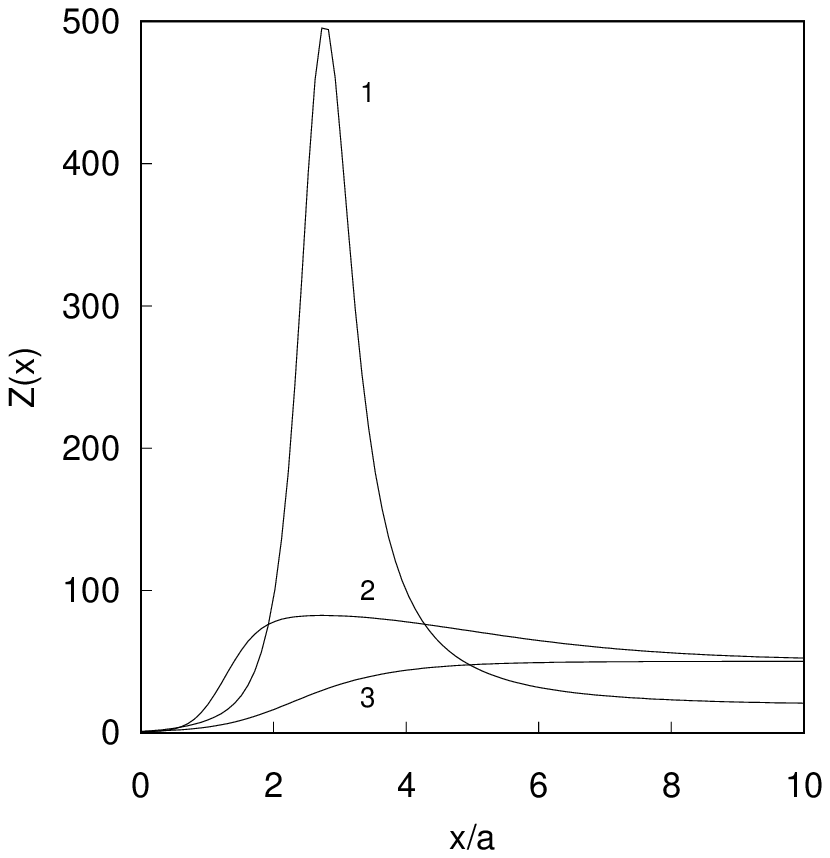}
\caption{Plot of $Z(x)$ for sDOS (1, $\varkappa=0.88$), eDOS (2, $\varkappa=0.99$), in both cases $c(x)=2\exp(-x/a)-\exp(-x/2a)$, and for eDOS for $c(x)=\exp(-x/a)$ (3, $\varkappa=0.99$).}
\label{Zmax}
\end{figure}

The reason for the decrease of the mobility and subsequent linear rise of $\ln v$ for the nonpositive $c(x)$ is the same for both the exponential and Gaussian DOS: the development of the maximum of $Z(x)$, while for the positive $c(x)$ the corresponding $Z(x)$ monotonously grows with $x$  (see Fig. \ref{Zmax} for the exponential DOS, the behavior of $Z(x)$ for the Gaussian DOS with the nonpositive correlation function is evident from \eq{1D_mu}). For the exponential DOS in the vicinity of the maximum $\ln Z$ may be approximated by the parabolic function similar to \eq{x_{min}}, thus giving the same linear dependence for the mobility.  For sDOS the maximum is more pronounced than for eDOS with the same nonpositive $c(x)$, thus the corresponding decrease of the mobility is greater, but in other respects the behavior is qualitatively similar. We may also note that for the particular power law correlation function used for calculation of the mobility field dependence in Figs. \ref{dec_mu} and \ref{Gauss} the minimal value of the correlation function $c_{min}=-0.2158..$ is rather low.

Decrease of the mobility with increasing field in weak field region for $c(x) < 0$ is different from the seemingly the same behavior previously observed in Monte Carlo simulation and initially attributed to the formation of field induced traps in materials with the  spatial disorder.\cite{Borsenberger:5447,Bassler:15}  As it was later successfully explained, the apparent decrease of the mobility is a direct consequence of the carrier diffusion giving the dominant contribution to the carrier transport at low fields, and the proper consideration of the diffusive contribution eliminates the effect.\cite{Hirao:1787,Hirao:4755,Cordes:094201}  In our case the decrease of the mobility is a real effect related to the particular behavior of $Z(x)$ for the nonpositive $c(x)$. This effect also differs significantly from the possible decrease of the mobility in strong field region which may be attributed to various and very different mechanisms such as the saturation of the drift carrier velocity for the Miller-Abrahams hopping rate \cite{Bassler:15} or the development of the inverted regime for the Marcus rate.\cite{Seki:14305,Fornari:9997}

A reasonable question is to what extent the major features of the mobility field dependence for the nonpositive correlation function remain the same in 3D case. We cannot provide a true reliable answer to this question, but the comparison with the results of the extensive computer simulation for 3D carrier transport for the Gaussian DOS give some hints (as it was already noted). For example, for the mobility at low field $E\rightarrow 0$ and strong disorder $\sigma/kT \gg 1$
\begin{equation}\label{lowE}
  \ln \mu \simeq -C_d \left(\frac{\sigma}{kT}\right)^2
\end{equation}
and the best analytic result using renormalization group approach gives for the constant $C_d=1/d$ irrespectively of the correlation properties of the random energy landscape (here $d$ is the dimension of the space).\cite{Deem:911} This result agrees well with the exact solution of the 1D case \cite{Dunlap:542,Parris:2803} and computer simulations for the 3D case.\cite{Novikov:4472,Novikov:954,Novikov:2532} We see that the dependence on $d$ is limited to the variation of the constant $C_d$. Mobility field dependence for moderate fields (\ref{1D_mu}) again agrees well with the results of 3D simulation for the dipolar disorder ($n=1$), quadrupolar disorder ($n=3$), and uncorrelated disorder ($n\rightarrow\infty$), and only the proportionality coefficient depends on $d$.\cite{Novikov:4472,Novikov:2584,Novikov:954,Novikov:2532}  Certainly, the comparison with the case of the Gaussian DOS are not sufficient: a thorough study of 3D simulation of the charge transport in the exponential DOS having nonpositive correlation function will be considered in a separate paper.

\section{Conclusion}

We considered the effect of the negative spatial correlation on the transport properties of a particle (charge carrier) moving in the random energy landscape having the exponential DOS. It was found that the general feature of the negative correlation is the decrease of the mobility with the field (bare velocity $v_0$)  in the low field region and subsequent development of the universal mobility field dependence $\ln v/v_0 \propto v_0$ for stronger field strengths.

For the case of symmetric DOS the correlation effect is even more dramatic, leading to the increase of the temperature $T_c$ of the transition to the transport regime where the average carrier velocity goes to $0$ for the infinite transport layer; for the positive correlation function the critical temperature is equal to $T_b=U_0/k$ and coincides with the temperature of the breakdown of the quasi-equilibrium stationary state. To the best of out knowledge this is the first case where the critical temperature $T_c$ for the exponential DOS depends on the correlation.

Regime of the zero average velocity (the dispersive transport) for the non-correlated eDOS takes place at $T < T_b$ and demonstrates current transients decaying as the power law of time
\begin{equation}
I(t) \propto \begin{cases} t^{-(1-\alpha)}, \hskip10pt t < t_0, \\ t^{-(1+\alpha)}, \hskip10pt t > t_0, \end{cases}
\label{i(t)}
\end{equation}
where $t_0$ is some transit time obtained by the analysis of transient in double logarithm coordinates $\ln I$ vs $\ln t$ and $\alpha=kT/U_0 < 1$.\cite{Silver:352,Schwarz:148,pope1999,Santos:8034}  Our result implicates that if $T_b < T_c$, then the transport regime for $T < T_c$ may be different from the usual dispersive regime developing for the non-correlated eDOS, and even if the transient demonstrate the same temporal dependence (\ref{i(t)}), then the relation between $\alpha$ and $U_0$ should be different. In addition, the dependence of the average carrier velocity on the transport layer thickness $L$  should be different from the usual form $v\propto 1/L^{1/\alpha-1}$.

We found also that for the positive correlation function $c(x)$ the general features of the transport (strong dependence of the mobility on the applied field in the close vicinity of $T_c$ and the functional form of the mobility field dependence) are the same as for the previously studied XY-model. Thus, we may be sure that these features reflect the true characteristics of the particle transport in the random correlated exponential landscape and not the particular way to generate such landscape.

Most interesting results are obtained for the symmetric DOS. Probably such DOS could be found in locally ordered amorphous organic semiconductors if the dominant contribution to the energy landscape is provided by quadrupoles and axises of neighbor quadrupoles tend to be oriented in parallel.\cite{Novikov:124711}  Details of the ordering are not important and the domains of approximately parallel quadrupoles having the linear size of $\simeq 5$ molecules clearly demonstrate exponential tails; in addition they have a tendency to form the ani-correlated distribution of energies (see Fig. 8 in the cited paper). Similar tails could be formed in the dipolar semiconductors but only under a more restricting condition: neighbor dipoles have to be oriented in the antiparallel fashion. A model of the approximately exponential sDOS was suggested also by Brown and Shaheen.\cite{Brown:135702}  Recently the shape of the DOS in several organic glasses serving as the popular charge transporting materials has been studied.\cite{Stankevych:44050}  In two materials with molecules having low dipole moments (i.e., the very type of the materials where we can expect the dominant quadrupolar contribution to the DOS) the exponential tails of the DOS have been observed, while for materials with molecules having high dipole moments the DOS has the Gaussian shape.

Some degree of the local order is natural for the arrangement of spacious and highly asymmetric organic molecules, thus the development of the exponential tails seems to be a common phenomenon. At the same time, the validity of the direct extrapolation of the one-dimensional consideration to the charge transport in real semiconductors is not obvious. More careful study of the of the charge transport in locally ordered organic semiconductors could probably shed more light on the effect of spatial correlation on the critical temperature of the breakdown of the normal transport regime with the nonzero average velocity.

We found that for the nonpositive correlation function the characteristic feature of the charge transport is the decrease of the mobility with increasing $E$ in the low field region followed by the raise $\ln\mu\propto E$. This behavior takes place even for the Gaussian DOS. Observation of the predicted mobility field dependence in some amorphous semiconductor could hint to the development of the negative correlation in that material.

Shift of the transition temperature $T_c$ and development of the specific mobility field dependences directly affect all processes and applications where charge transport in amorphous semiconductors with the exponential DOS having negative correlation is important. Possible examples are operation of  electronic devices, kinetics of electrochemical reactions, and charge carrier recombination.

\section*{Acknowledgements}
Financial support from the Ministry of Science and Higher Education of the Russian Federation (A.N. Frumkin Institute) and Program of Basic Research of the National Research University Higher School of Economics is gratefully acknowledged.





%


\end{document}